\pdfoutput=1 
\documentclass{sig-alternate}
\newfont{\mycrnotice}{ptmr8t at 7pt}
\newfont{\myconfname}{ptmri8t at 7pt}
\let\crnotice\mycrnotice%
\let\confname\myconfname%

\permission{ }
\conferenceinfo{Balisage: The Markup Conference.}{August 2014,  Washington, DC, USA.}
\copyrightetc{ }
\crdata{ http://dx.doi.org/10.1042/...}

\usepackage{balance}
\usepackage{algorithm,algorithmicx,algpseudocode}

\newdef{myaxiom}{Axiom}
\newdef{mylemma}{Lemma}
\newdef{myprop}{Property}
\newdef{myexample}{Example}
\newdef{definition}{Definition}

\begin{document}

\title{Non-hierarchical Structures: How to Model and Index Overlaps?}
\numberofauthors{2}
\author{
\alignauthor
Faegheh Hasibi \\
       \affaddr{Norwegian University of Science and Technology}\\
       \email{faeghehh@idi.ntnu.no}
\alignauthor
Svein Erik Bratsberg\\
       \affaddr{Norwegian University of Science and Technology}\\
       \email{sveinbra@idi.ntnu.no}
}

\maketitle

\begin{abstract}
Overlap is a common phenomenon seen when structural components of a digital object are neither disjoint nor nested inside each other. Overlapping components resist reduction to a structural hierarchy, and tree-based indexing and query processing techniques cannot be used for them. Our solution to this data modeling problem is TGSA (Tree-like Graph for Structural Annotations), a novel extension of the XML data model for non-hierarchical structures. We introduce an algorithm for constructing TGSA from annotated documents; the algorithm can efficiently process non-hierarchical structures and is associated with formal proofs, ensuring that transformation of the document to the data model is valid. To enable high performance query analysis in large data repositories, we further introduce an extension of XML pre-post indexing for non-hierarchical structures, which can process both reachability and overlapping relationships.

%

\end{abstract}

\section{Introduction}

Texts, data, objects (digital or otherwise) can be viewed from different perspectives. No perspective depends upon another and each exists separate from the other. By attempting to encode different perspectives that are independent of each other, one can encounter ``overlapping elements''. The most common example is a document with two distinct structural views, when the logical view is \textit{section/subsection/ paragraph} and the physical view is \textit{page/column}. Each single structural view of this document is a hierarchy and the components are either disjoint or nested inside each other. The overlapping issue arises when one structural element cannot be neatly nested into others. For instance, when a paragraph starts in one page and terminates in the next page. Similar situations can appear in videos and other multimedia contents, where temporal or spatial constituents of a media file may overlap each other \cite{Salembier:2007kx}.

The most used model for expressing structure of documents is based on hierarchies, which ensures that each region is nested within another and the regions can be accessed by use of parent-child or ancestor-descendant relationships. This tree data structure requires organizing structural information of digital objects in a single tree, which is not applicable for overlapping structures. In other words, tree-based markup languages -such as XML- and the corresponding indexing and retrieval techniques are not sufficient for documents with non-hierarchical structures~\cite{Hasibi:2013:IQO}.

Scholars have introduced several solutions for overlapping problem. TEI (Text Encoding Initiative) consortium \cite{Consortium:2014pi} suggests several methods to deal with non-hierarchical structures in SGML or XML context. However, these methods are  just syntactical solutions to represent non-hierarchical structures and non of them are based on a well-defined data model. On the other hand, there are some non-XML markup languages such as SGML CONCUR \cite{Sperberg-McQueen:1999mi}, LMNL \cite{Piez:uq} and TexMECS \cite{Huitfeldt:2001ys}, in which  TexMECS is based on a data well-defined model called GODDAG.


In order to efficiently query over overlapping structures in large data repositories, the structural elements of an overlapping data model should be efficiently summarized and indexed. To this end, we consider documents with continuous, ordered structural elements (as common features of non-hierarchical structures) and introduce a data model together with an XML-compatible parsing and indexing method. In other words the scope of this paper is overlap-only marked up documents, such that the order of appearance of document elements is unique. Overlap-only documents are also addressed in \cite{Marcoux:2008sp} and \cite{Marcoux:2013eu}, where the authors mostly focused on serializability of the data model. 

The contributions of this paper are three-fold: 1) We formally define TGSA\footnote{TGSA pronounced as ``tegsa".} (Tree-like Graph for Structural Annotations) as a new data model for overlapping structures, which preserves simplicity while being expressive. 2) We introduce an efficient algorithm to map annotated documents to the TGSA data model. Our algorithm  is compatible with XML parsing algorithms  and can efficiently generate overlapping data model (TGSA) comparing to \cite{Sperberg-McQueen:2004ys}. 3) We extend XML pre-post indexing method to the TGSA data model, which can represent parent-child and ancestor-descendant relationships as well as overlaps.  

This paper is organized as follows. Section 2 motivates this research by providing use cases and applications of non-hierarchical structures. Section 3 presents the background and related works. Modelling and indexing methods are represented in Section 4, where subsection 4.1, 4.2 and 4.3 describe the TGSA data model, the TGSA construction algorithm and indexing methods, respectively. Section 5 concludes the paper.

\section{Use Cases and Applications}

\begin{figure*}
\centering
\includegraphics[width=\textwidth]{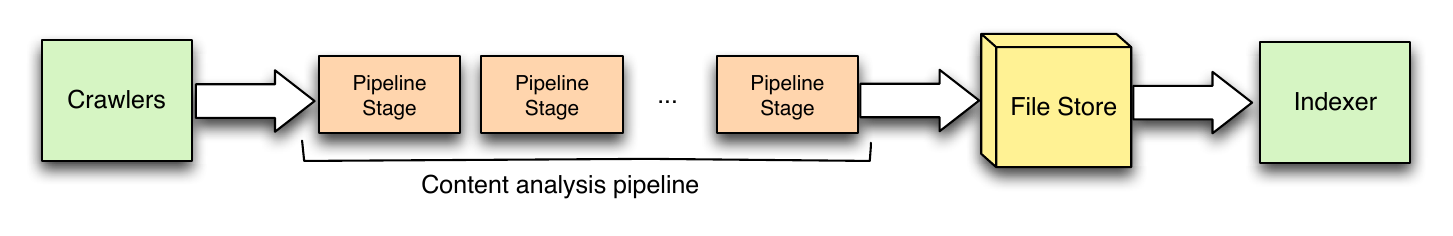}
\caption{Data processing pipeline in large scale search engines}
\label{pipe}
\end{figure*}

Overlapping is a situation that is more common than it may be thought of.  In general, annotating several aspects of an object (i.e. stream of data) implies overlapping structures. Lots of scholars encounter overlaps in the area of computational linguistic, speech and complex text analysis. Some of these situations are as follows: 

\paragraph*{Content analysis of textual data} Many search applications use a pipeline process for analysing input data. For instance, FSIS (FAST search for internet sites), which is a Microsoft search platform, uses such pipeline to detect content of unstructured or semi-structured documents \cite{ESP:2008fk, FSIS}. As shown in Figure \ref{pipe}, this pipeline consists of several components, where each component analyses one aspect of data and adds annotations to the input. These annotations can be structural information such as sentences, paragraphs and links or extracted entities such as addresses, locations and names. Since each component extracts annotations independent of other components, some annotations may overlap each other.

\paragraph*{Question Answering systems} QA systems need to search among a large collection of annotated documents to find the answer for a natural language query. These systems, mostly have access to a rich linguistic and semantic annotation of data, where overlapping is very likely to happen. As it is discussed in \cite{ogilvie2010retrieval, Bilotti:2007fk}, in order to get effective results for complex queries,  a QA system needs to handle overlaps for both indexing and query processing.

\paragraph*{Annotating non-textual objects} Overlapping structures can be seen not only in documents, but also while encoding other kinds of digital data. For instance, to classify a movie database based on genres, the time interval of each genre should be annotated by movie analysers. Among these annotations, there might be scenes that are annotated for more than one genres. Example 1 illustrates this scenario, where \textit{``dream"} scene overlaps both \textit{``romance"} and \textit{``musical"} genres. Handling overlapping annotations, allows classical structural information retrieval systems to answer the queries like: 

\begin{itemize}
\item Find all scenes that only contain romance content (i.e. no overlaps with other genres).
\item Find all scenes that have both musical and romance contents (i.e. scenes that overlap two genres).
\end{itemize}

\begin{myexample}
\begingroup
    \fontsize{8pt}{10pt}\selectfont
\begin{verbatim}

<sample>
  <video>
    <scene id="love" s="00:00" e="00:38"/>
    <scene id="dream" s="00:39" e="01:44"/>
  </video>
  <genres>
    <genre id="romance" s="00:00" e="00:45"/>
    <genre id="musical" s="00:46" e="01:34"/>
    <genre id="romance" s="01:00" e="01:44"/>
  </genres>
</sample>
\end{verbatim}
\endgroup
\end{myexample}

As shown in Example 1, video annotations are encoded using stand-off annotations, where the original data is separated from structural views. Stand-off annotation is a natural technique to encode annotations of non-textual data and also linguistic annotations of text documents \cite{Alink:2006ij, ogilvie2010retrieval}.

\paragraph*{Change tracking of documents}
Text change tracking is a feature provided by text editors such as Microsoft Office to allow users to revise a document. Tracking of text changes is a known challenging topic, since this information needs to be stored in a markup language and the new structure of text may conflict with the former one. In other words, whenever a change affects the structure of a text (such as merging two paragraphs or splitting a paragraph), the notations of this change break the hierarchical structure of current text encoding. Example 2 demonstrates a case where the text change overlaps existing structure of document \cite{Di-Iorio:2011pb}.

\begin{myexample}
\begingroup
    \fontsize{8pt}{10pt}\selectfont
\begin{verbatim}

<p>The end of one paragraph 
  <change-start change-id= "1"/> </p> 
<p>and <change-end change-id="1"/> 
  the start of another paragraph</p>
\end{verbatim}
\endgroup
\end{myexample}

\paragraph*{Digitizing old manuscripts} This task is another domain that needs dealing with non-hierarchical structures. Building electronic version of old manuscripts requires encoding massive amounts of information, such as textual content, physical location of texts, linguistic information, visibility of characters and information about damages of manuscript \cite{Dekhtyar:2005uq}.

\section{Background and Related Work}

\subsection{Overlapping Data Models}
The difficulty of handling overlaps is that overlapping structures are not hierarchies and the popular markup languages, such as XML are based on hierarchies. As a result of lacking an adequate overlapping data model, several data structures have been proposed to describe overlapping structures. In the following, we discuss these data structures and their abilities to model different types of overlaps. 
	
\paragraph*{Multiple Hierarchies (SGML CONCUR)}
The most straightforward model for the overlapping problem is to keep multiple hierarchies in a single document. This model is captured by the CONCUR feature of SGML, which maintains multiple structural views of a document. It actually extends the SGML/XML data model to a model, where multiple trees (with the same frontier) can be encoded within a single document.

The CONCUR model is represented as a part of SGML and consequently it is a legitimate and maintainable approach for overlapping problem. However, this model is not widely implemented as a solution of overlaps. Here is a list of CONCUR drawbacks \cite{Sperberg-McQueen:2004ys, DeRose:2005dz, Barnard:1988vn}:
\begin{enumerate}
\item The model is not able to constrain relations among DTDs. For example a quote element may overlap paragraph elements but not chapter elements in the other DTD \cite{DeRose:2005dz}.
\item CONCUR does not provide self-overlaps, where two elements, with the same name, coinciding each other. In such cases one element have to be moved to another hierarchy, which means CONCUR should support unpredictable DTDs.
\item Concur cannot simply model the deletion, insertion, duplication, or reordering of data in the various views \cite{Sperberg-McQueen:2004ys}.
\end{enumerate}

\paragraph*{MCT}
The Multi-Coloured Tree (MCT) \cite{Jagadish:2004zl} data structure was introduced  for storing multiple hierarchies. In MCT, different coloured hierarchies are built on top nodes, where they can share the same nodes.   As discussed in \cite{Iacob:2005uq}, MCTs are not suitable for representing multiple hierarchies in document-centric XML. One important issue is that there is no global order for children in different hierarchies.	

\paragraph*{LMNL}
LMNL (Layered Markup and Annotation Language) is a data model associated with a markup language \cite{Piez:uq, Piez:2012fk}. LMNL data model is based on layers rather than hierarchies.It represents documents without forcing elements into a hierarchy and can capture overlaps.

\paragraph*{GODDAGs}
 GODDAG (Generalized Ordered Descendant Directed Acyclic Graph) \cite{Sperberg-McQueen:2004ys} is a well-defined data model, which is introduced  to represent documents with overlapping structures. Principally, GODDAG is a  directed acyclic graph (DAG), where each non-terminal node has ordered descendants. 
 
 GODDAG has two variations: restricted and generalized GODDAG.  Restricted GODDAG is capable of representing overlaps; however its constraints rules out the possibility of modelling non-contiguous elements. Here are three constraints of restricted GODDAG, which are relaxed in generalized GODDAG: 1) Leaf nodes are ordered. 2) Each non-terminal dominates a contiguous subsequence of leaves. 3) No two nodes dominate the same subsequence of the frontier. By relaxing these constraints, generalized GODDAG can represent discontinuous elements. 
 
 Sperberg-McQueen et al. introduced and algorithm to convert MECS encoded documents to restricted GODDAG \cite{Sperberg-McQueen:2004ys}, while no algorithms is suggested for generalized GODDAG. In this paper, we show that the TGSA data model and its corresponding algorithm outperforms GODDAGs in  efficiency.
 
\subsection{Querying of Overlapping Structures}
Querying over non-hierarchical structures requires a mechanism that relates structural regions to each other. Iacob et al. \cite{iacob2005queries} extended XPath as EXPath  to query overlaps over GODDAG structures. XIRAF \cite{Alink:2006ij} is another system that allows querying over overlapping annotations by moving from one hierarchy to another hierarchy. XIRAF's query approach is based on Burkowski's \cite{Burkowski:1992fy} work, which adds four new overlapping operations to XPath queries.

It should be noted that all of these query approaches are developed for domain specific applications. However, query processing for large data repositories needs investigations on indexing structures.

\subsection{Structural Indexing}
Overlapping data structures can be modelled by either graphs or tree-like structures, such as GODDAGs. To the best of our knowledge, there is no research directly investigating on the overlapping indexing, however there has been large number of studies on XML and graph indexing.

According to \cite{Gou:2007vn}, there are two main classes of structural indexing for XML data: numbering schemes and index graph schemes.  The former is used for path joining, while the latter is for path selection in answering XML queries. Zhang et al. \cite{Zhang:2001fk} proposed a numbering scheme for XML documents, called \textit{PrePost} encoding. This model is capable of processing parent-child as well as ancestor-descendant relationships. \textit{Dewey coding} \cite{Tatarinov:2002ve} is another famous numbering scheme, which can be maintained easier than PrePost method. Jin \cite{Jin:2009zr} introduced a 3-hop indexing scheme, which is targeted for directed graphs with high edge-vertex density. 

\section{Modelling and Indexing}
In this section, we divide the problem of handling  non-hierarchical structures into three issues and we dig into each issue in the following subsections. The first issue is which data model can represent hierarchies as well as overlaps?  Although graphs can represent any relationships, we need less generalized but expressive enough data model that can represent overlapping structures efficiently. The second is how to  parse and convert encoded documents to the data model?  and the third is how to index these structures to efficiently process structural queries? 
%

\subsection{TGSA: Tree-like Graph for Structural Annotations} 
\begin{figure*}
\centering
\includegraphics[scale=0.3,width=\textwidth]{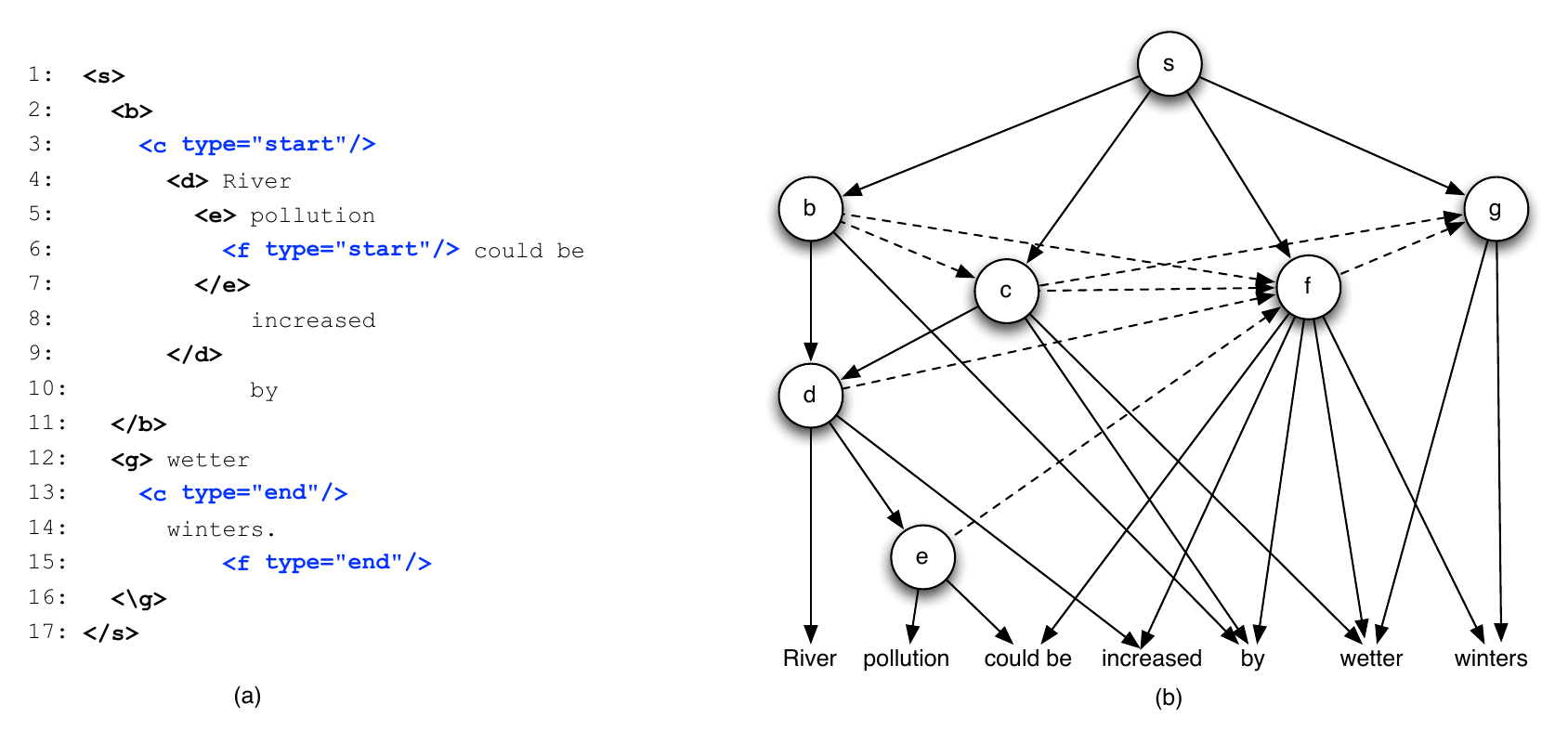}
\caption{An example of the TGSA data model. Element \textit{``c"} overlaps elements \textit{``b, f"}, and \textit{``g"}, while precedes elements \textit{``f"} and \textit{``g"}. Element \textit{``f"} overlaps elements \textit{``b,c,d,e,g"}, while it precedes element \textit{``g"}. 
(a) XML encoded document. The \textit{start} and \textit{end} of overlapping elements are indicated with the \textit{type} attribute.
(b)The TGSA data model. Solid arcs and dashed arcs represent \textit{``P-C"} (parent-child) and \textit{``O"} (overlapping) relationships, respectively. }
\label{tgsaEx}
\end{figure*}

In general, structural annotations of documents are represented with graphs. However, for well-formed XML documents, these graphs are basically tree data structures, where edges represent parent-child relationships that are  explicitly encoded in the document.  In order to extend this data model to capture overlapping annotations, we introduce a new type of edge that represents overlapping relationship. In addition,  we allow nodes to have more than one parent. As a result, the new data model would be a graph that satisfies some properties of trees. We call this new data model TGSA, which stands for Tree-like Graph for Structural Annotations.

TGSA is basically a directed graph that has a set of nodes and arcs. The nodes correspond to the structural elements or text-nodes.  Each arc in TGSA graph is labelled with either ``P-C" (parent-child) or ``O" (overlapping).  Examples of the data model are shown in Figure \ref{tgsaEx} and Figure \ref{tgsaCons} (d), in which overlapping and parent-child relationships are shown using dashed and solid arcs, respectively. We consider TGSA as a two relational directed graph, defined as follows:

\begin{definition} 
Two-relational graph $G = (V, E)$ has a set of vertices $V$ and a set of arcs $E : V \times V$  denoted by triples of form $(v_{i}, l,  v_{j})$, where $l \in \{``P-C", ``O"\}$. 
\end{definition}

Based on this definition, we define TGSA as follows:

\begin{definition}
Let $G = (V, E)$ be a two-relational directed graph. G is TGSA \textit{iff}:
\begin{enumerate}
\item G is connected and has no loops.
\item All modes have \textbf{atleast} one incoming edge except one root node that has no incoming edge.
\item Given nodes $v_{i}$ and $v_{j}$, $e= (v_{i}, ``P-C",  v_{j})$ \textit{iff} $v_{i}$ is parent of  $v_{j}$.
\item Given nodes $v_{i}$ and $v_{j}$, $e=(v_{i}, ``O",  v_{j})$ \textit{iff} $v_{i}$ precedes $v_{j}$ AND $v_{i}$ and $v_{j}$ are overlapping.
\item Given nodes $v_{i}$ and $v_{j}$, if there is an indirect path from  $v_{i}$ to $v_{j}$, then the arc $e=(v_{i}, ``P-C",  v_{j})$ cannot exist.
\end{enumerate}
\end{definition}

Paths in  TGSA are defined whenever two nodes are connected via $``P-C"$ arcs. In other words,  $``O"$ arcs do not create paths between nodes.

\begin{definition}
A path in TGSA is a sequence of $\{x_{1},x_{1}, ..., x_{n}\}$ such that  $(x_{1}, ``P-C",  x_{2})$ , $(x_{2}, ``P-C",  x_{3})$ , ..., $(x_{n-1}, ``P-C",  x_{n})$ are graph arcs of the type $``P-C"$ and the $x_{i}$ are distinct.
\end{definition}

Paths in TGSA are transitive relations, in which given a path from $u$ to $v$ and $v$ to $w$, there is always a path from $u$ to $w$. Unlike paths, overlapping relations are not transitive.

\begin{myaxiom}
Given a TGSA graph $G = (V, E)$ and arcs $e1 = (v_{i}, ``O",  v_{j})$ and $e2 =(v_{i}, ``O",  w_{j})$, \textbf{does not} imply that $v_{i}$ ``overlaps" $w_{j}$.
\end{myaxiom}

Based on the above definitions, TGSA can represent elements that overlap each other in very complex ways.  
Moreover, TGSA does not constrain relationships between annotations across the document; except that each annotation is related to a continuous piece of text and the document elements are ordered.  

\paragraph{TGSA vs. GODDAG}
TGSA is similar to restricted GODDAG in three aspects: 1) Cycles are not allowed in both data models. 2) The constraint of being only one path between two nodes is relaxed in both of them, which is one of the main properties of trees. This implies that each node can have multiple parents and in TGSA, each node can overlaps multiple nodes. 3)  Each node dominates or overlaps a contiguous set of nodes, which means discontinuous elements can not be modelled.

On the other hand, unlike restricted GODDAG, TGSA  mark overlapping elements using a different arcs type. By introducing \textit{overlapping} arcs, TGSA can represent order of overlapping nodes, which is not possible in restricted GODDAG. 
TGSA can represent continuous, ordered elements, while GODDAGs can handle disordered elements (e.g. two parents with the same order but in different order). Such constraints together with introducing \textit{overlapping} relations, makes TGSA a data model that be efficiently indexed, queried and converted from structural annotations.


\subsection{Constructing TGSA from Annotated Documents}

We now present an algorithm for building TGSA from annotated text documents, as shown in Algorithm \ref{alg1}. The input to the algorithm is an in-line tagged document, where start and end of each node is labelled in the text.  Such documents ensure that the document elements are ordered. The algorithm is independent of encoding format of input, in which any encoding method or markup language can be used to represent overlaps. The only requirement is that, the start and end delimiters should be specified among the text.  It must be stated that stand-off annotations, where the start and end of nodes are specified by positional values are not supported by this algorithm. However, if the elements of a document are continuous and ordered, they can be converted to TGSA, even if they are encoded by stand-off annotations. For the sake of simplicity, we consider in-line tagged documents as the input of Algorithm \ref{alg1} and show how document elements are mapped to TGSA.

\begin{algorithm}                      
\caption{TGSA Construction}          
\label{alg1}                           
\begin{algorithmic}[1]
\Require Annotated document $D$
\Ensure TGSA graph $G=(V,E)$
\end{algorithmic}
  \bigskip

\textbf{begin}
\begin{algorithmic}[1]
    	\State $V = \emptyset$ , $E = \emptyset$
    	\State $L \leftarrow$ Initialize a list for open nodes
    	\While{$\neg EOF(D)$}
    		\State $u \leftarrow$ read next physical unit
    		\If{$u$ is text node}
    			\State $E.add(\{L.latestEntry, ``P-C", u\})$
    		\ElsIf{$u$ is opening tag for node $n$}
    			\State $V.add(n)$
  				\State $E.add(\{L.latestEntry, ``P-C", n\})$
 			 	\State $D \leftarrow$ initialize a set   
    			\State $n.D = \emptyset$     \Comment{ To save overlapped descendants}
    			\State $L.add(n)$ 
    		\ElsIf{$u$ is end tag for node $n$}
    			\If{$L.latestEntry = n$}
    				\State $L.remove(n)$
    			\Else \Comment{n overlaps some nodes}
     				\State $x \leftarrow  L.firstNodeAddedAfter(n)$
     				\State $E.remove(\{n, ``P-C", x\})$ 
     				\State $E.add(\{n.parent, ``P-C", x\})$
					\For{all node $y$ added to $L$ after $n$ }
						\State $E.add(\{n,``O", y\})$
						\State $addParentChildRelation(n,y)$
					\EndFor
    			\EndIf    				
    		\EndIf
     	\EndWhile
\end{algorithmic}
\textbf{end}
\bigskip

\begin{algorithmic}[1]
\Function{addParentChildRelation}{n,y}
	\For{node $c$ in $y.children()$}
		\If{$c \notin n.D$} \Comment $c$ is not descendant of $n$
			\State $E.add(\{n, ``P-C", c\})$
			\For{all ancestor $a$ of $n$ in $L$}
				\If{$c \notin$ nodes added to $L$ after $n$}
					\State $a.L.add(c)$
				\EndIf
			\EndFor
		\EndIf
	\EndFor
\EndFunction
\end{algorithmic}
\end{algorithm}

\begin{figure*}
\centering
\includegraphics[width=\textwidth]{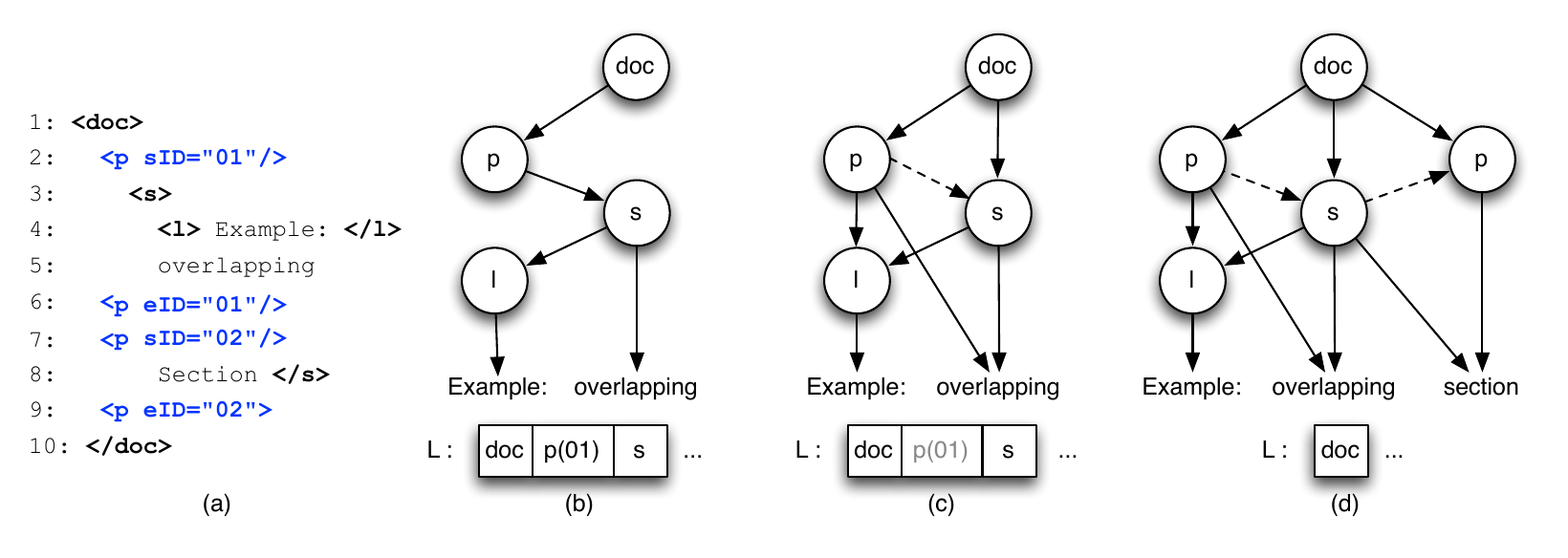}
\caption{Constructing the TGSA data model from a non-hierarchical annotated document. $L$ represents the list of open nodes in Algorithm \ref{alg1}. (a) XML encoded document. Attributes \textit{sID} and \textit{eID} indicate the start and end of each node. (b) Parsing line 5. (c) Parsing line 6. (d) Parsing line 10.}
\label{tgsaCons}
\end{figure*}

To construct the TGSA graph $G=(V,E)$, we need to initialize a list called $L$ to keep track of open elements. A  node will be added to and removed from this list when an open and end tag is seen, respectively. As shown in Figure \ref{tgsaCons} (b), we first assume that the document is hierarchical and every two adjacent open nodes have parent-child relationship. Whenever an end tag is seen, we need to check weather the node is overlapping with other nodes or not. If the node is the latest element added to the list $L$, then it is not overlapping, otherwise the node is overlapping  with the nodes added after it. The situation is demonstrated in Figure \ref{tgsaCons} (c), where node $p$ overlaps node $s$. In such cases, four changes should be applied to the TGSA graph: 1) Remove ``P-C" arcs between $p$ and $s$. 2) Add ``O" arcs from $p$ to $s$. 3) Define a new parent for node $s$ by adding ``P-C" arcs from parent of node $p$ to node $s$. 4) Define new children for node $p$ by adding ``P-C" relation from node $p$ to the children of node $s$.

One of the challenges of the TGSA construction algorithm is when a new node should be added as a child of an overlapping node (the last step in the above paragraph). In this step, we need to be sure adding this relationship is legal and satisfies the last item of Definition 2.

To clarify, let us consider two overlapping nodes $d$ and $f$ in Figure \ref{tgsaEx}. When the algorithm reaches the end tag of node $d$, children of node $f$ are ``could be" and ``increased". Here we should only add ``increased" as a child of node $d$, since there is a path from $d$ to ``could be" through node $e$. The straightforward solution to overcome this problem is to check all possible paths from node $d$ to all children of node $f$ and add ones that are not descendants of node $d$. This solution is very costly and will increase time complexity of the algorithm drastically.

We efficiently address this issue by introducing a new set for each node in the list of open nodes $L$. This set keeps the descendants of each node that are assigned to the node due to overlapping relations. For each node in the list $L$, we initialize this empty set, which is called $D$ in algorithm \ref{alg1}. Function $AddParentChildRelation$ describes how we use this set to add new children to each overlapping node. Given node $n$ and node $c$ as a child of node $y$, the function first checks whether $c$ is member of $n.D$ or not. If not, a new arc will be added from node $n$ to $c$. In addition, node $c$ will be added to the $D$ set of all nodes added before $n$ to the list $L$. The following proof describes how this solution will guarantee that every node will be added to the TGSA graph according to its definition.

\begin{mylemma} 
Given TGSA graph $G= (V, E)$ and a list of open elements $L = \{a, b, ..., d, e, ..., n, ...\}$, if node $e$ overlaps node $n$ and $x$ is a child of node $n$, then there is an indirect path from each precedents of node $e$ in the list $L$ to node $x$.
\end{mylemma}

\begin{proof}
To prove that there exists indirect path between two nodes, we need to proves that two nodes have ancestor-descendant relationship (and not parent-child relationship).

Node $e$ overlaps node $n$ implies that node $d$ is parent of node $e$. Therefore, node $d$ is ancestor of node $x$.

All other precedents of node $e$ in list $L$ are either ancestor/parent of node $d$ or overlaps node $d$. Assuming node $b$ as one of precedents of node $d$: 
\begin{enumerate}
	\item If node $b$ is ancestor/parent of node $d$, then node $b$ is ancestor of node $x$. 
	\item If node $b$ overlaps node $e$, then children of node $d$ are assigned as children of node $b$. Since node $d$ is ancestor of node $x$, then node $b$ is also ancestor of node $x$. 
\end{enumerate}
\end{proof}

The TGSA construction algorithm is a very efficient way of processing and representing overlapping structures, which is also compatible to hierarchical structures. Given $n$ as number of nodes in an annotated document, time complexity of this algorithm for hierarchical structures is $O(n)$. For non-hierarchical structures, we do not consider lookup costs of function \textit{addParentChildRelation}, since the algorithms only needs to test that a node is not definitely in a set and this can be efficiently done by bloom filters \cite{Bloom:1970fk}. 

Comparing the TGSA algorithm to the algorithm represented in \cite{Sperberg-McQueen:2004ys}, our algorithm has less time complexity than GODDAG construction algorithm. GODDAG algorithm needs to remove illegal arcs from GODDAG after the whole graph is build, which means the algorithm requires to find all possible paths in the graph, and remove redundant ones. Since there are multiple paths between two nodes, this process has large affect on the performance of algorithm.

\subsection{Indexing Method for TGSA }

Indexing is an essential method for providing high-performance query processing for large data repositories. Like tree-based data, non-hierarchical data needs efficient methods of storage to  summarize the structural view of a document. Without indexing, one should traverse the data model forward and backward to find structural relationships.   

Motivated by this, indexing method for non-hierarchical structures can be inspired by XML indexing methods. XML indexes include value and structural indexes, where the former is to index string literals and the later is to index structure of documents. In this paper, we extend XML pre-post indexing to non-hierarchical structures, as it provides more efficient support for finding relationships between nodes.

\begin{figure*}
\centering
\includegraphics[width=\textwidth]{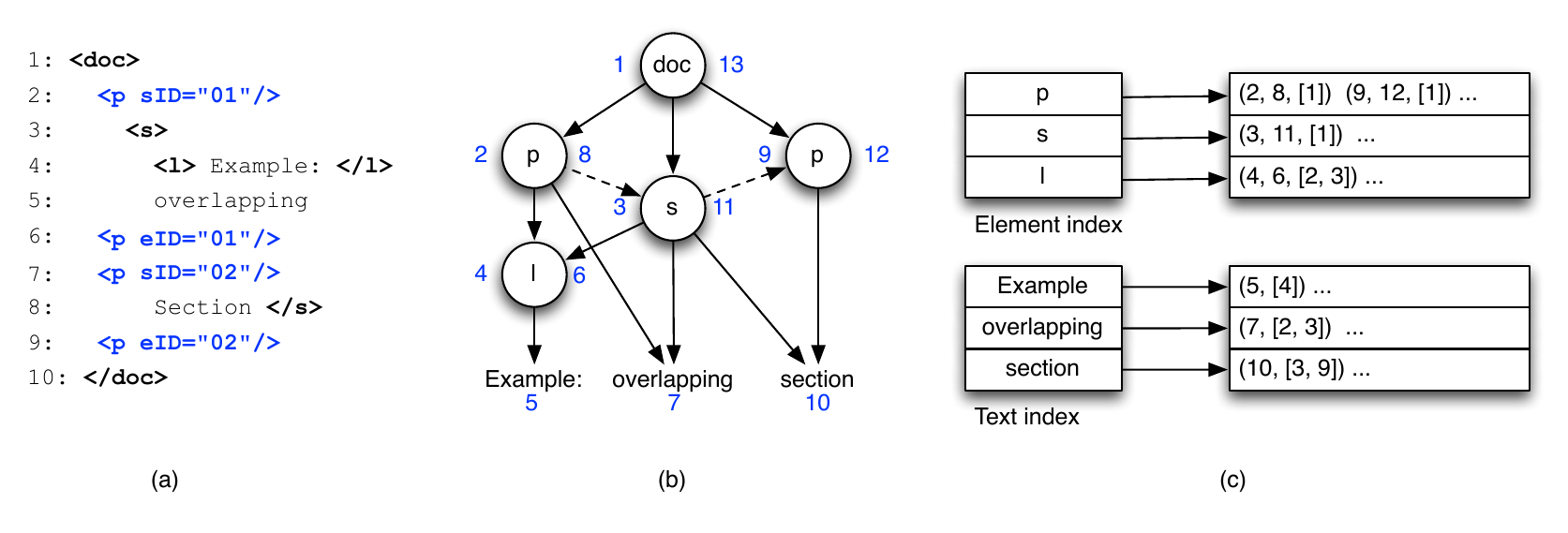}
\caption{Constructing the TGSA data model and inverted index from a non-hierarchical annotated document. (a) XML encoded document. Attributes \textit{sID} and \textit{eID} indicate the start and end of each node. (b) TGSA data model. The numbers at the left and right side of each node indicate the position of start and end tags of the element, respectively. (c) Inverted list of text index. (d) Inverted list of element index.}
\label{tgsaIndex}
\end{figure*}

Pre-post indexing labels each node with two numbers, $(start, end)$, which corresponds to the position of start and end tag of each element. By the help of $(start, end)$, we can easily find the ancestor-descendant and overlapping relationships relationships. However, this information is not enough to determine parent-child relationships. This relationship can be supported by only adding parent(s) of each to the $(start, end)$ pair. Figure \ref{tgsaIndex} illustrates an example of pre-post indexing for non-hierarchical structures. It should be stated that unlike XML, each entry in the structural index can have multiple parents, which is the intrinsic feature of TGSA. For hierarchical structures, this indexing method only contains one parent for each node. The following properties show how different relationships are supported by this indexing method.

\begin{myprop} (Overlapping relationship) 
Node $a$ precedes and overlaps node $b$ \textit{iff} $ a.start < b.start$ AND $a.end < b.end$
\end{myprop}

\begin{myprop} (Ancestor-Descendant relationship) 
Node $a$ is ancestor of node $b$ \textit{iff} $a.start < b.start$ AND $b.end<a.end$
\end{myprop}

\begin{myprop} (Parent-Child relationship) 
Node $a$ is parent of node $b$ \textit{iff} $b.parent=a.start$
\end{myprop}

Interval indexing requires much less storage space than storing the whole TGSA data model. Moreover, it provides very efficient support of different relationships by few numerical comparison. On the other hand, the drawback of this indexing method is the difficulty in maintaining the index, since updating the data will result major changes in the $start$ and $end$ position of each node.

\section{Conclusion and Future Work}
While XML is the dominant format for storing structural information, studies show that there are a series of structural data that cannot be presented and stored in hierarchical formats. In order to query these non-hierarchical structures, the first need is an expressive and simple data model that can represent different structural relationships.

In this paper, we considered documents with continuous, ordered structural elements, and we formally introduced a  novel data model called TGSA, which can handle overlapping and parent-child relationships. The main features of this data model are its compatibility with hierarchical data and handling overlapping elements. Converting structural elements to the TGSA data model is the second step on handling overlapping structures. To this end, we introduced an efficient algorithm for constructing TGSA, which  is validated with formal  proofs.  We also proposed an extension of XML pre-post indexing method to address high performance query analysis for large data repositories. This indexing method can represent parent-child, ancestor descendant and overlapping relationships between two nodes.

The focus of this paper was on theoretical aspects of non-hierarchical structures. The on-going research is likely to focus on experimental aspects of these structures using real world data sets and to adapt hierarchical query processing methods to non-hierarchical structures. 

Another directions of research we hope to pursue in near future is investigating on serializability between TGSA and annotated documents and supporting stand-off annotated documents in the TGSA construction algorithm.
\balance
\providecommand{\bibfont}{\small}

\raggedright
\bibliographystyle{abbrv}
{\raggedright \small
\bibliography{sr-hasibi}
}

\end{document}